\documentclass[10pt, conference, letterpaper]{IEEEtran}
\usepackage{float}
\usepackage{amsmath}
\usepackage{amssymb}
\usepackage{amsthm}
\usepackage{multirow}
\usepackage{tabularx}
\usepackage{graphicx}
\usepackage{subfigure}
\usepackage{authblk}
\usepackage[linesnumbered, ruled]{algorithm2e}

\usepackage{threeparttable}
\usepackage{slashbox}
\usepackage{algorithmic}
\usepackage{tabularx}
\usepackage{booktabs}
\usepackage{amsfonts}
\usepackage{dsfont}
\usepackage[all]{xy}

\def\squarebox#1{\hbox to #1{\hfill\vbox to #1{\vfill}}}

\begin{document}

\title{Founding Digital Currency on \\Imprecise  Commodity}

\author{Zimu Yuan$^{\dagger\ddagger}$, and Zhiwei Xu$^{\ddagger}$  \\ 
$^{\dagger}$University of Chinese Academy of Sciences, China\\
$^{\ddagger}$Institute of Computing Technology, Chinese Academy of Sciences, China\\
\{yuanzimu, zxu\}@ict.ac.cn
}

\maketitle \thispagestyle{empty}

\begin{abstract}
Current digital currency schemes provide instantaneous exchange on precise commodity, in which "precise" means a buyer can possibly verify the function of the commodity without error. However, imprecise commodities, e.g. statistical data, with error existing are abundant in digital world. Existing digital currency schemes do not offer a mechanism to help the buyer for payment decision on precision of commodity, which may lead the buyer to a dilemma between having to buy and being unconfident. In this paper, we design a currency schemes IDCS for imprecise digital commodity. IDCS completes a trade in three stages of handshake between a buyer and providers. We present an IDCS prototype implementation that assigns weights on the trustworthy of the providers, and calculates a confidence level for the buyer to decide the quality of a imprecise commodity. In experiment, we characterize the performance of IDCS prototype under varying impact factors.
\end{abstract}




\section{Introduction}
Digital currency plays an important role on today's online transaction processing. In past years, research and business community have proposed many digital currency schemes, e.g., \cite{scheme1}\cite{scheme2}\cite{scheme3}\cite{scheme4}, for various application scenarios. Those schemes provide instantaneous transaction and ownership transfer on precise commodity, in which "precise" means perfect conformity to fact or truth that a buyer can possibly verify the function of the commodity without error. Meanwhile, there exists uncountably imprecise data with error existing in digital world. For instance, statistics, e.g., Gross Domestic Product (GDP), usually collected from multiple sources, are easily influenced by negligence, limited manpower or even falsification. However, existing digital currency schemes do not have a mechanism to help the buyer to judge and pay on the precision of data. As a consequence, when taking the imprecise data as digital commodity for trade, the buyer may not have enough information to determine if the source or provider is trustworthy (or which provider is more trustworthy), and thus has no idea on the precision of the data he needs to buy. Once the buyer has paid for erroneous data, he can hardy get the cash back. To help the trade of imprecise commodity proceeding normally, it is desired to give support to the trading process in scheme level.

To realize a digital currency scheme on imprecise commodity, we have found the challenges lie in establishing the trustworthy relationship between buyer and providers. Consider that a trading process is happening. Initially, the buyer does not know anything about the imprecise commodity and providers, and he would need additional information as decision aid. The difficulty in scheme design is the scheme also starts with zero knowledge on the commodity and the providers. Without prior knowledge, the scheme can hardy help the buyer to rate the trustworthiness of the providers as well as to find the truth out of the commodity's views given by the providers. Then, another challenge as the trading process goes is when the buyer still holds doubtful attitude before payment, how can the scheme help the buyer to prompt his confidence on choice; otherwise, the buyer may be caught in a dilemma between having to buy and being unconfident. Finally, once the buyer decides to pay for a provided view of the commodity believed to be the truth, it is also important for the scheme to determine a fair distribution of payment to the providers while preventing the buyer from being deceived by malicious providers.

In this paper, we set out to tackle the challenges in realizing digital currency scheme on imprecise commodity. In particular, we are interested in paying for the truth while identifying malicious providers that may probably disrupt currency exchange system from the buyer's side. As a summary, this paper makes the following contributions:
\begin{itemize}
  \item We design a \textbf{C}urrency \textbf{S}cheme for \textbf{I}mprecise \textbf{D}igital commodity (IDCS) that completes a trade through three-stage processing between a buyer and providers.
  \item We present an IDCS prototype implementation that assigns weights to the providers according to their trustworthy, and calculates a confidence level for the buyer to decide the quality of a commodity view.
  \item We experiment under varying impact factors, which characterizes the performance of our IDCS prototype.
\end{itemize}

The remainder of this paper is organized as follows. We introduce the preliminaries in Section \ref{preliminary}. We formally present IDCS in Section \ref{IDCS_Scheme}. We describe an IDCS prototype implementation in Section \ref{IDCS_Prototype}. We provide experimental results in Section \ref{IDCS_Experiment}. We review related work in Section \ref{related_work}. Finally, we conclude our paper in Section \ref{conclusion}.

\section{Preliminaries} \label{preliminary}
This section outlines the preliminaries of IDCS, including adversary model, weight model, and payment model. All of these models exhibit the relationship between buyer and providers.

\subsection{Adversary Model} \label{adversary}
We assume the malicious providers behave in a Byzantine manner, i.e., supplying the imprecise view of digital commodity arbitrarily. The providers are independent of each other, i.e., they do not collude to supply a same bias view of a commodity. (Otherwise, the providers can be divided into independent groups such that any two providers belong to different groups give unrelated views. Here we omit the dependent case in model assumption.) In addition, a provider may also supply a imprecise view due to negligence or limited manpower, besides out of malicious intent.

We assume both the buyer and the ledger server that acts as an intermediary between the buyer and the providers are honest. The buyer indeed wants to purchase the commodity, and does not purposely return the commodity back. The ledger server processes the trade faithfully following the specification of IDCS. The case of dishonest buyer or ledger server is quite different in assumption with the honest case, which needs a entirely different scheme on trade. In this paper, we simplify the assumption, and choose to purely focus on the digital currency from the buyer's honest side.

\subsection{Weight Model}
IDCS weights the reliability of the providers. The weights are derived based on the evaluation of trading commodities. At the beginning, IDCS starts with zero weights on the providers. As trades happen, IDCS incrementally adjusts weights according to the evaluation results on commodities supplied by the providers. The underlying principle is more reliable providers supply more trustworthy views on a commodity, and thus should be assigned with higher weights. Here we consider the single truth scenario, i.e., although the providers supply different views of a commodity, there is only one truth view. Suppose that there are total $m$ providers supplying their views $v_1,v_2,...,v_m$ on a commodity $v$ separately. The weight model calculates a estimated truth view $v^*$ by the following equation:
\begin{equation} \label{e_truth}
v^{*}=\frac{\sum_{i=1}^{m}w_i v_i}{\sum_{i=1}^{m}w_i}
\end{equation}
where $w_i$, $i=1,2,...,m$, are the weights for the providers. $v^*$ is averaged by adopting provider weights as view weights. In Section \ref{implemented_method}, we will present an implementation of weight model in our IDCS prototype.

\subsection{Payment Model}
All the trades are proceeded through a ledger server. The process of a trade is started from the buyer. When a buyer wants to purchase a commodity, he firstly specifies a payment mode and registers it on the ledger server. Then if a provider accepts the payment mode, the provider can supply a view of the commodity to the ledger server and apply for payment. Finally, the trade is done after the buyer confirms the payment for the supplied view on the ledger server. Suppose that there are total $m$ providers supplying their views $v_1,v_2,...,v_m$ on a commodity $v$ separately. Formally, we define the payment function as follow:
\begin{equation}
[c_1,c_2,...,c_m]=p(C,W,V)
\end{equation}
where $W=\{w_1,w_2,...,w_m\}$ includes the weights assigned to views, $V=\{v_1,v_2,...,v_m\}$ and $c_1,c_2,...,c_m$, $\sum_{i=1}^{m} c_i=C$, are the distribution of currency to the $m$ providers respectively. In Section \ref{implemented_example}, we introduce three payment functions for evaluation of our IDCS prototype.

\section{Imprecise Digital Currency Scheme (IDCS)} \label{IDCS_Scheme}
\begin{figure}
\centering
\includegraphics[width=3.4in]{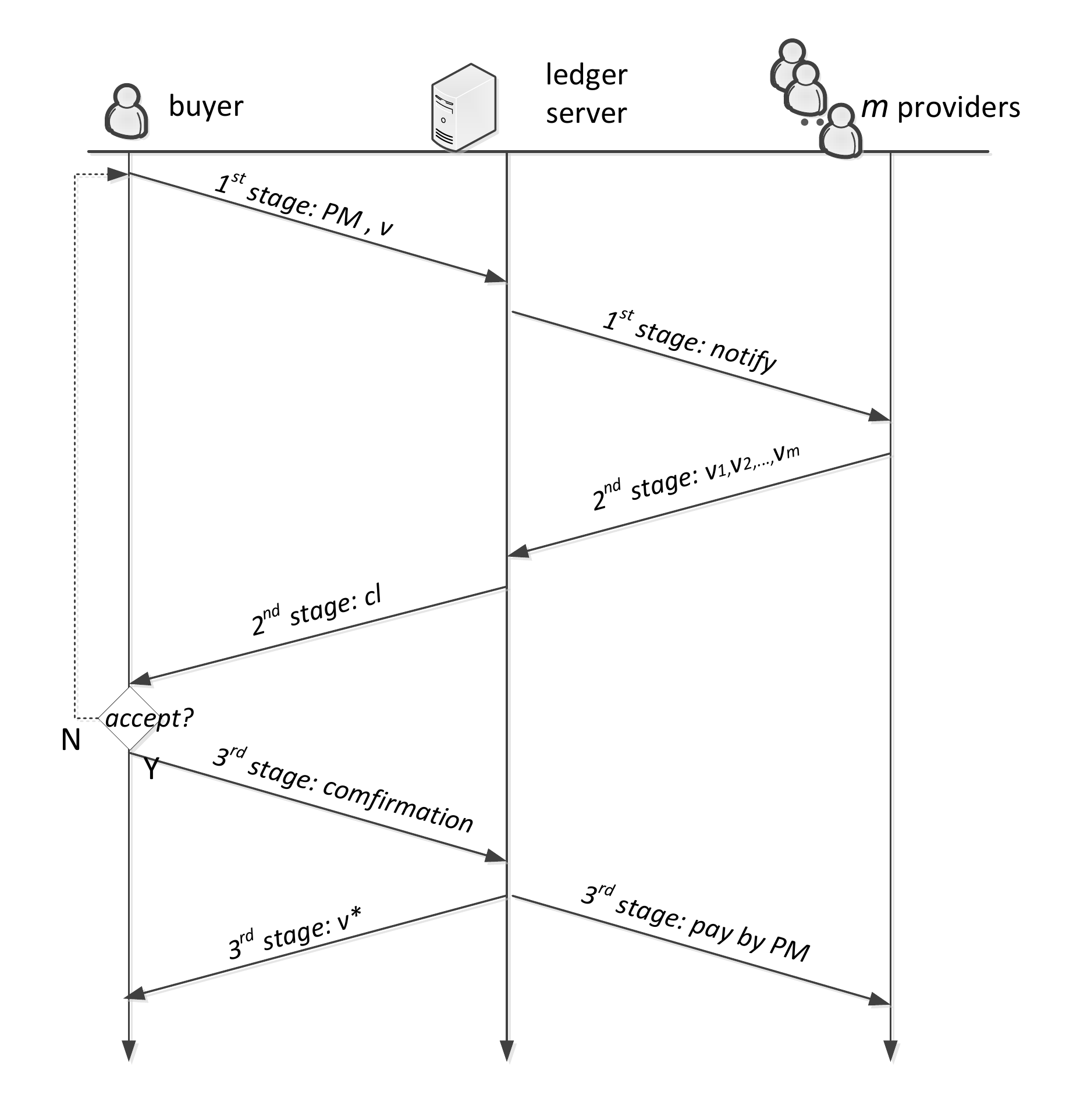}\\
\caption{\textrm{IDCS completes a trade in three stages. In the 1st stage, the buyer declares payment mode $PM$ for a commodity $v$; then the ledger server notifies (or posts) it to the providers. In the 2nd stage, a total of $m$ providers supply their views $v_1,v_2,...,v_m$ for $v$; then the ledger server estimates a truth view $v^*$, calculates a confidence level $cl$ as an evaluation of quality on $v^*$, and send $cl$ to the buyer. In the 3rd stage, If the buyer admits the confidence level $cl$, he sends the confirmation back to the ledger server; on receiving the confirmation, the ledger server sends the estimated truth view $v^*$ to the buyer and completes payment by $PM$. Else if the buyer is not satisfied with the confidence level $cl$, he is allowed to redeclare $PM$ by changing the payment functions and increasing payment, and the trading process returns to the 1st stage.}} \label{fig_IDCS}
\end{figure}

IDCS completes a trade through three-stage processing between the buyer and the providers (Figure \ref{fig_IDCS}). All of the three stages use a ledger server as trusted intermediary. We describe them in details as follows:
\begin{itemize}
\item 1st stage: the buyer declares payment mode $PM$ for a commodity $v$:
    \begin{equation} \label{e_PM}
    PM(v)=
    \begin{cases}
    \displaystyle p_1(C_1,W,V) & \qquad cl \leq cl_1  \\
    \displaystyle p_2(C_2,W,V) & \qquad cl_1 < cl \leq cl_2   \\
    \displaystyle ...... \\
    \displaystyle p_n(C_n,W,V) & \qquad cl > cl_{n-1}\\
    \end{cases}
    \end{equation}
    where $cl_1$, $cl_2$, ..., $cl_{n-1}$ are the confidence levels that mark off different payment functions, i.e., given $p(0,W,V)$, $cl \leq 0.5$, the buyer will not pay if the confidence level is below or equal to $0.5$, whereas given $p(100,W,V)$, $cl > 0.5$, the buyer pay $100$ currency units according the specification of function $p$ and weights $W$ if the confidence level is greater than $0.5$. The buyer registers $PM$ on the ledger server. Then, the ledger server notifies or posts $PM$ to the providers.
\item 2nd stage: If a provider accepts the payment mode $PM$, he could supply his view of the commodity to the ledger server. Suppose there are total $m$ providers that supply their views $v_1,v_2,...,v_m$. The ledger server assigns weights $w_1,w_2,...,w_m$ according the reliability of the providers, and estimates a truth view $v^*$ (Equation (\ref{e_truth})) based on the weight assignment. Then, the ledger server evaluates the quality of view $v^*$ by calculating a confidence level $cl$, and send it to the buyer for confirmation.
\item 3rd stage: If the buyer admits the confidence level $cl$, he sends the confirmation back to the ledger server. The ledger server sends the estimated truth view $v^*$ (Equation (\ref{e_truth})) to the buyer, and completes the payment to the providers by the payment mode $PM$. Else if the buyer is not satisfied with the confidence level $cl$, he is allowed to redeclare $PM$ by changing the payment functions and increasing payment, i.e., $\forall cl_{i-1}<cl\leq cl_i$, the redeclared payment function $p'_i(C'_i,W,V)$ has $C'_i \geq C_i$. With redeclaration, the trading process returns to the 1st stage.
\end{itemize}

\section{IDCS Prototype Implementation} \label{IDCS_Prototype}
As can be seen in Section \ref{IDCS_Scheme}, the process of IDCS relies on the weight assignment and evaluation on the providers. In this section, we present an implemented prototype of IDCS that assigns weights to the providers based on the quality of the views their supply and calculate the confidence level for the buyer. Then, we introduce a practical case on IDCS prototype.

\subsection{An Implemented Method on Weight Assignment} \label{implemented_method}
The weights are assigned according to the error of the supplied views. When a provider just joins in, the implemented system of IDCS has no knowledge on the provider. The ledger server initially starts several trades on imprecise commodities, the ground truth view of which are already known, with the provider. Then, IDCS can assesses the reliability of the providers by deriving the mean error $\mu$ and error variance $\sigma^2$ on the views they supply. Let $S$ denote the set of views corresponding to the series of trades started by the ledger server. For a provider, we have:
\begin{equation}
\mu = \frac{\sum_{s \in S} d(s^g-s)}{|V|}
\end{equation}
\begin{equation}
\sigma^2 = \frac{\sum_{s\in S} (\mu-s)^2}{|V|}
\end{equation}
where $d(s^g-s)$ is the distance between the ground truth $s^g$ and the view $s$ supplied by the provider. Then, as the provider joins in subsequent practical trades, IDCS can incrementally adjust $\mu$ and $\sigma^2$ according to the quality of views on commodities supplied by the provider.

We assume that providers are independent of each other (otherwise they can divided into independent groups) in the adversary model (Section \ref{adversary}). Thereupon, we can use Gaussian distribution to describe the error on the views of commodities supplied by the providers. Suppose that there are total $m$ providers. For the $i$th provider, we have:
\begin{equation}
e_i \sim G(\mu_i, \sigma_i^2)
\end{equation}

For a commodity $v$, IDCS applies the weighted averaging strategy (Equation (\ref{e_truth})) to calculate a estimated truth view $v^*$. With the assumption that the providers are independent of each other, we have the error $e^*$ of view $v^*$ following Gaussian distribution:
\begin{equation}
e^* \sim G(\frac{\sum_{i=1}^{m}w_i \mu_i}{\sum_{i=1}^{m}w_i}, \frac{\sum_{i=1}^{m}w^2_i \sigma^2_i}{(\sum_{i=1}^{m}w_i)^2})
\end{equation}
Without loss of generality, we restrict $\sum_{i=1}^{m}w_i=1$. Suppose that we have an error threshold value $e_T$. The objective of weight assignment in IDCS is to maximize the probability $P(|e^*|<e_T)$:
\renewcommand{\arraystretch}{1.5}
\begin{equation} \label{e_optimal}
    \begin{array}{ll}
        \max \; {P(|e^*|<e_T)}\\
        s.t. \; \sum_{i=1}^{m}w_i=1, w_i \geq 0.
    \end{array}
\end{equation}
\renewcommand{\arraystretch}{0.667}

\begin{figure}
\centering
\includegraphics[width=1.8in]{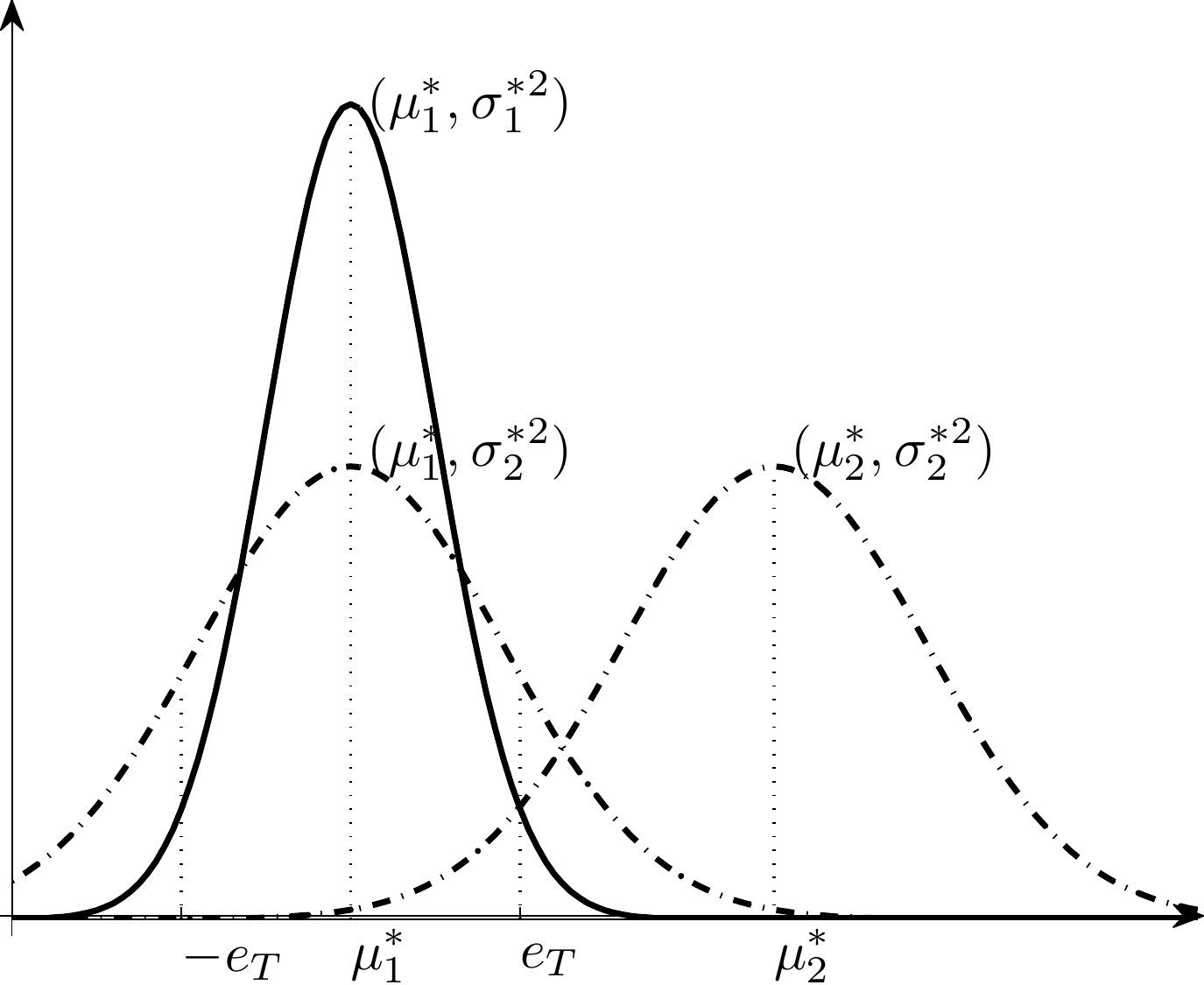}\\
\caption{\textrm{An example on Gaussian distribution.}} \label{fig_gauss}
\end{figure}

Let $\mu^*=\frac{\sum_{i=1}^{m}w_i \mu_i}{\sum_{i=1}^{m}w_i}$ and $\sigma^*=\frac{\sum_{i=1}^{m}w^2_i \sigma^2_i}{(\sum_{i=1}^{m}w_i)^2}$. In Figure \ref{fig_gauss}, we have $\mu^*_1<\mu^*_2$ and $\sigma^*_1<\sigma^*_2$. As can be seen, under the Gaussian distribution $G(\mu^*_1, \sigma^{*2}_1)$, $P(|e^*|<e_T)$ has a greater value compared with the other two distributions. In other words, to maximize $P(|e^*|<e_T)$, the views should be weighted with the smallest combination of $|\mu^*|$ and $\sigma^*$. Substituting Gaussian probability-density function in $P(|e^*|<e_T)$, we have:
\begin{equation} \label{e_prob}
P(|e^*|<e_T)=\int_{-e_T}^{e_T} \frac{1}{\sqrt{2\pi}\sigma^*} Exp(-\frac{(e^*-\mu^*)}{2\sigma^{*2}}) \; d\,e^*
\end{equation}
Unfortunately, it can be verified that the objective of maximizing $P(|e^*|<e_T)$ with Equation (\ref{e_prob}) cannot be directly solved. As a consequence, we are unable to derive the smallest combination of $\mu^*$ and $\sigma^*$ straightly.

Consider that if the error $e_i$ of a view $v_i$ is more likely between the interval $[-e_T,e_T]$ (namely a greater value of $P(|e_i|<e_T)$), then the view $v_i$ is more trustworthy. We apply an approximate weight assignment:
\begin{equation}
w_i \varpropto P(|e_i|<e_T)
\end{equation}
Normalize the weight assignment with the constraint of $\sum_{i=1}^{m}w_i=1$. For $e^*$, its weight assignment is:
\begin{equation}
w_i=\frac{P(|e_i|<e_T)}{\sum_{j=1}^{m} P(|e_j|<e_T)}
\end{equation}
We set the confidence level $cl=P(|e^*|<e_T)$ for IDCS, which means that the error $e^*$ of the estimated view $v^*$ has the probability of $cl$ between interval $[-e_T,e_T]$.

\subsection{GDP Dataset Case on IDCS Prototype} \label{implemented_example}
Here we introduce a case of Gross Domestic Product (GDP) dataset on our IDCS prototype. Usually, GDP statistics are collected from multiple sources (or to say providers), and their precisions are easily influenced by external force, e.g. limited manpower, falsification.

\subsubsection{The GDP Dataset} \label{GDP_dataset_case}
We choose the GDP statistics of China in 1994 - 2014 \cite{GDP} as example. The GDP statistics contain 3 independent parts, statistics by the expenditure approach, income approach, and productive approach respectively. These statistics are listed in Table \ref{table_GDP_growth_rate}. We have the following relationship of statistics:
\begin{equation} \label{e_GDP}
\begin{cases}
\displaystyle GDP\_EA=FCE+GCF+NE \\
\displaystyle GDP\_IA=NPT+WC+DFA+BB \\
\displaystyle GDP\_PA=FI+SI+TI \\
\end{cases}
\end{equation}
All of these statistics can be provided as indicators for GDP growth. We take them as views of GDP growth rate (from 1995 to 2014). Since we do not actually know the most trustworthy view, we choose GDP\_PA as the ground truth. The mean error and standard deviation of the other views to GDP\_PA are also listed in Table \ref{table_GDP_growth_rate}. Note that the statistics of GDP\_FA, FCE, GCF, and NE in 2014, the statistics of GDP\_IA, NPT, WC, DFA and BB in 2004, 2008, and 2011-2014 are not given in the official site of National Bureau of Statistics \cite{GDP}. We use the GDP growth rate of their previous year to fill these blanks.

\subsubsection{The Reference Weight Methods} We implement four other weight methods besides IDCS weight (IDCSW) method (Section \ref{implemented_method}) as references for evaluation of IDCS prototype. Two of these methods are Mean and Median:
\begin{itemize}
  \item Mean: assign weight 1 to the views $v_1,v_2,...,v_m$ to estimate $v^*$ (Equation (\ref{e_truth})).
  \item Median: assign weight 1 to the view of median value (when $m\%2=1$) or the two views of median value (when $m\%2=0$), and weight 0 to the other views to estimate $v^*$.
\end{itemize}
Also, we can find a great lot of truth-finding methods \cite{ra_1}\cite{ra_2}\cite{rb_b1}\cite{ra_3}\cite{rb_a1} in research literatures. The basic idea behind them can be mainly concluded into two categories: finding the most likely value by manipulating the data itself, and calculating the final result based on the reliability of sources. Thus, we implement two methods, K-voting and K-sources, in spirit related to the two categories respectively.
\begin{itemize}
  \item K-voting: Let any two views $v_i$ and $v_j$ vote their distance $d(v_i,v_j)$ to each other, and assign weight 1 to the $k$ views nearest to all the other views and weight 0 to the other views to estimate $v^*$.
  \item K-sources: Based on prior information, assign weight 1 to the $k$ most trustworthy views and weight 0 to the other views to estimate $v^*$.
\end{itemize}
In our GDP dataset case, the distance $d(v_i,v_j)$ is equal to the difference of GDP growth rate between $v_i$ and $v_j$. For the K-voting and K-sources method, we set $k=3$, and let the prior information of K-sources method be the randomly selected 10 years' statistics by default. For IDCSW method, we initially starts 10 trade for each of 10 years' statistics as well, and set $e_T=1$ in confidence level $cl$ calculation.

\begin{table}
\renewcommand{\arraystretch}{1.0}
\caption{Views on GDP growth rate} \label{table_GDP_growth_rate}
\centering
\begin{tabular}{p{1.2cm}p{3.5cm}p{1cm}p{1.2cm}}
  \hline
  Views & Full Name & Mean Error & Standard Deviation \\
  \hline
  FCE & Final Consumption Expenditure & 2.4069 & 1.5291 \\
  GCF & Gross Capital Formation & 3.8193 & 2.9389 \\
  NE & Net Exports & 33.6287 & 34.5794 \\
  GDP\_EA & GDP by The Expenditure Approach & 1.2462 & 0.9685 \\
  NPT & Net Production Tax & 3.9390 & 3.4461 \\
  WC & Worker Compensation & 4.1153 & 5.3371 \\
  DFA & Depreciation of Fixed Assets & 3.6984 & 2.3672 \\
  BB & Business Balance & 10.7253 & 14.3010 \\
  GDP\_IA & GDP by The Income Approach & 3.0893 & 3.6595 \\
  FI & GDP of The First Industry & 4.8382 & 3.2961 \\
  SI & GDP of The Secondary Industry & 1.6570 & 1.1663 \\
  TI & GDP of The Tertiary Industry & 2.6926 & 1.9201 \\
  GDP\_PA & GDP by The Productive Approach & 0 & 0 \\
  \hline
\end{tabular}
\end{table}

\begin{table}[!t]
\renewcommand{\arraystretch}{1.05}
\caption{Error payment of GDP dataset case} \label{table_error_1}
\centering
\begin{tabular}{cccc}
  \hline
  Error Payment & $(1,1)$ & $(3,\frac{1}{d})$ & $(m,\frac{1}{d^2})$\\
  \hline
  IDCSW &    1.5000  &  1.0295 &   1.0678 \\
  Mean &    1.7000  &  1.5963 &   1.4956 \\
  Median &    1.5000  &  1.1242 &   1.2180 \\
  3-voting &    1.8000  &  1.3826 &   1.5112 \\
  3-sources &    1.9000  &  1.7080 &   1.5810 \\
  \hline
\end{tabular}
\end{table}

\subsubsection{The Payment Functions} We implement three payment functions for evaluation of weight methods in error distribution of currency. The three payment functions are different in the number of providers who receive payment. Suppose that $<v_{(1)},v_{(2)},...,v_{(m)}>$ is the sorted sequence in ascending order according to the distance $d(v_{(i)},v^*)$, $i=1,2,...,m$.
\begin{itemize}
  \item $(1,1)$: the provider of view $v_{(1)}$ gets all the $C$ currency units.
  \item $(3,\frac{1}{d})$: the provider of view $v_{(1)}$, $v_{(2)}$ and $v_{(3)}$ get $c_{(i)}=\frac{C\cdot \prod_{k=1,k\neq i}^3 d(v_{(k)},v^*)}{\sum_{j=1}^{3} \prod_{k=1,k\neq j}^3 d(v_{(k)},v^*)}$, $i=1,2,3$, currency units respectively.
  \item $(m,\frac{1}{d^2})$: the provider of view $v_{(1)},v_{(2)},...,v_{(m)}$ get $c_{(i)}=\frac{C\cdot \prod_{k=1,k\neq i}^m d^2(v_{(i)},v^*)}{\sum_{j=1}^{m} \prod_{k=1,k\neq j}^m d^2(v_{(j)},v^*)}$, $i=1,2,...,m$, currency units respectively.
\end{itemize}

\begin{table*}[!t]
\renewcommand{\arraystretch}{1.05}
\caption{Error payment under varying malicious providers $mp$} \label{table_mp}
\centering
\begin{tabular}{ccccc|ccccc}
  \hline
  \multirow{5}{*}{$mp=3$} & Method & $(1,1)$ & $(3,\frac{1}{d})$ & $(m,\frac{1}{d^2})$ & \multirow{5}{*}{$mp=6$} & Method & $(1,1)$ & $(3,\frac{1}{d})$ & $(m,\frac{1}{d^2})$ \\
  \hline
  & IDCSW & 1.6000  &  1.2054  &  1.3383 &  & IDCSW & 1.7000  &  1.4981  &  1.4661 \\
  & Mean & 1.7000  &  1.4684  &  1.5294 &  & Mean & 1.8000  &  1.6265  &  1.5913 \\
  & Median & 1.6000  &  1.3427  &  1.4037 &  & Median & 1.8000  &  1.5767  &  1.5448 \\
  & 3-voting & 1.8000  &  1.5941  &  1.6209 &  & 3-voting & 2.0000  &  1.5120  &  1.5776 \\
  & 3-sources & 2.0000  &  1.5788  &  1.5771 &  & 3-sources & 1.8000  &  1.6108  &  1.5099 \\
  \hline
  \multirow{5}{*}{$mp=9$} & Method & $(1,1)$ & $(3,\frac{1}{d})$ & $(m,\frac{1}{d^2})$ & \multirow{5}{*}{$mp=12$} & Method & $(1,1)$ & $(3,\frac{1}{d})$ & $(m,\frac{1}{d^2})$ \\
  \hline
  & IDCSW & 1.6000  &  1.5485  &  1.5022 &  & IDCSW & 1.7000  &  1.6188  &  1.5636 \\
  & Mean & 1.7000  &  1.6183  &  1.5265 &  & Mean & 1.7000  &  1.6271  &  1.5643 \\
  & Median & 1.9000  &  1.7862  &  1.6927 &  & Median & 1.9000  &  1.8202  &  1.6296 \\
  & 3-voting & 1.9000  &  1.7734  &  1.6985 &  & 3-voting & 2.0000  &  1.9759  &  1.7985 \\
  & 3-sources & 1.8000  &  1.5427  &  1.5065 &  & 3-sources & 1.9000  &  1.4977  &  1.5042 \\
  \hline
\end{tabular}
\end{table*}

\begin{table*}[!t]
\renewcommand{\arraystretch}{1.05}
\caption{Error payment under varying manipulation factor $mf$} \label{table_mf}
\centering
\begin{tabular}{ccccc|ccccc}
  \hline
  \multirow{5}{*}{$mf=1.4$} & Method & $(1,1)$ & $(3,\frac{1}{d})$ & $(m,\frac{1}{d^2})$ & \multirow{5}{*}{$mf=1.6$} & Method & $(1,1)$ & $(3,\frac{1}{d})$ & $(m,\frac{1}{d^2})$ \\
  \hline
  & IDCSW & 1.5000  &  1.4231  &  1.4747 &  & IDCSW & 1.8000  &  1.6008  &  1.5711 \\
  & Mean & 1.7000  &  1.6946  &  1.6824 &  & Mean & 1.8000  &  1.7164  &  1.6230 \\
  & Median & 2.0000  &  1.7839  &  1.7773 &  & Median & 2.0000  &  1.9067  &  1.6787 \\
  & 3-voting & 2.0000  &  1.8414  &  1.8102 &  & 3-voting & 2.0000  &  1.8704  &  1.7098 \\
  & 3-sources & 1.7000  &  1.5274  &  1.5093 &  & 3-sources & 1.8000  &  1.6648  &  1.5864 \\
  \hline
\end{tabular}
\end{table*}

\subsubsection{The Evaluation Results}
To describe the quality of results, we define the error payment $e_{PM}$:
\begin{equation}
e_{PM}=\sum_{i=1}^m |c_{(i)}-c_{(i)^g}|
\end{equation}
where $c_{(i)}$ and $c_{(i)^g}$ are the amount of currency distributed to the $i$th provider by the payment function using the distance $d(v_{(i)},v^*)$ and $d(v_{(i)},v^g)$. (Recall that $v^g$ is the ground truth view.)

Let the trading commodity $v$ be the average GDP growth rate from 1995 to 2014. All the 12 statistics of GDP growth rate in Table \ref{table_GDP_growth_rate} are the providers that supply their views of commodity $v$. Assume the buyer accepts the results (or to say the confidence level) calculated by the weight methods, and confirms the payment $C=1$ with the function $(1,1)$, $(3,\frac{1}{d})$ and $(m,\frac{1}{d^2})$ respectively. We list the error payment of IDCS under weight method IDCSW, Mean, Median, 3-voting and 3-sources in Table \ref{table_error_1}. As can be seen, IDCSW has the least error payment among the weight methods. However, the result given by IDCSW also has a low confidence level $cl=22.30\%$ ($cl=P(|e^*|<e_T)$, Section \ref{implemented_method}), which means that the quality of the GDP statistics supplied by the providers should be improved.

\section{Experiment} \label{IDCS_Experiment}
To characterize the performance of IDCS, we manipulate the GDP dataset under three varying factors:
\begin{itemize}
  \item Malicious Provider $mp$: the number of malicious providers that manipulate the supplied views to prevent the buyer from finding the truth;
  \item Manipulation Factor $mf$: the degree of multiplying factor to the original view, i.e., $v'_i=mf\cdot v_i$;
  \item Improvement Factor $if$: the degree of accuracy improvement with per currency unit, using the following equation
      \begin{equation} \label{e_if}
        \frac{|v_i(j+1)-v^g|}{|v_i(j)-v^g|}=1-a\cdot e^{if\cdot(j+1)}
      \end{equation}
      where the $i$th provider with the payment $j+1$ currency unit can supply the view $1-a\cdot e^{if\cdot(j+1)}$ approaching to the ground truth view $v^g$ compared with the payment $j$ currency unit.
\end{itemize}
The Equation (\ref{e_if}) is raised with the intuition that the accuracy improvement by means of incentivization is at the fastest rate initially, and then slows down with more payment. In Experiment, we set $a=0.1$. The other experimental settings follow the setup in Section \ref{GDP_dataset_case} by default.

\subsection{Varying Malicious Providers} \label{experiment_mp}
Here the experiments are conducted under varying malicious providers $mp=3$, $6$, $9$ and $12$. The manipulation factor $mf$ is set to $1.2$. In experiment, the malicious views are randomly selected by 10 times for each $mp=3,6,9$. The error payments of function $(1,1)$, $(3,\frac{1}{d})$ and $(m,\frac{1}{d^2})$ with $C=1$ are shown in Table \ref{table_mp}. From the comparison results, we can see that (1) the error payment grows as the malicious providers increase; (2) the error payment decreases as the number of providers who receive payment increase; (3) IDCSW perform best among these methods (except for the $mp=12$ case that all the providers are with malicious intent), since IDCSW considers both mean and variance of error in its design. Besides, when changing the manipulation factor $mf$ under malicious providers $mp=3,6,9,12$, we can get similar results.

\subsection{Varying Manipulation Factor} \label{experiment_mf}
In this sub-section, we experiment on varying manipulation factor $mf=1.4$ and $1.6$. (The case $mf=1.2$ can be seen in previous sub-section.) The malicious providers $mp$, randomly selected by 10 times, is set to $6$. The error payments of function $(1,1)$, $(3,\frac{1}{d})$ and $(m,\frac{1}{d^2})$ with $C=1$ are shown in Table \ref{table_mf}. It can be seen that (1) the error grows in greater rate when manipulation factor increases compared with the change of malicious providers; (2) the error payment decreases as the number of providers who receive payment increase; (3) IDCSW has stable performance with the least error compared with other methods.

\subsection{Varying Improvement Factor} \label{experiment_if}
\begin{figure} \centering
\subfigure[malicious providers $mp=3$] { \label{fig_if:a}
\centering
\includegraphics[width=1.6in]{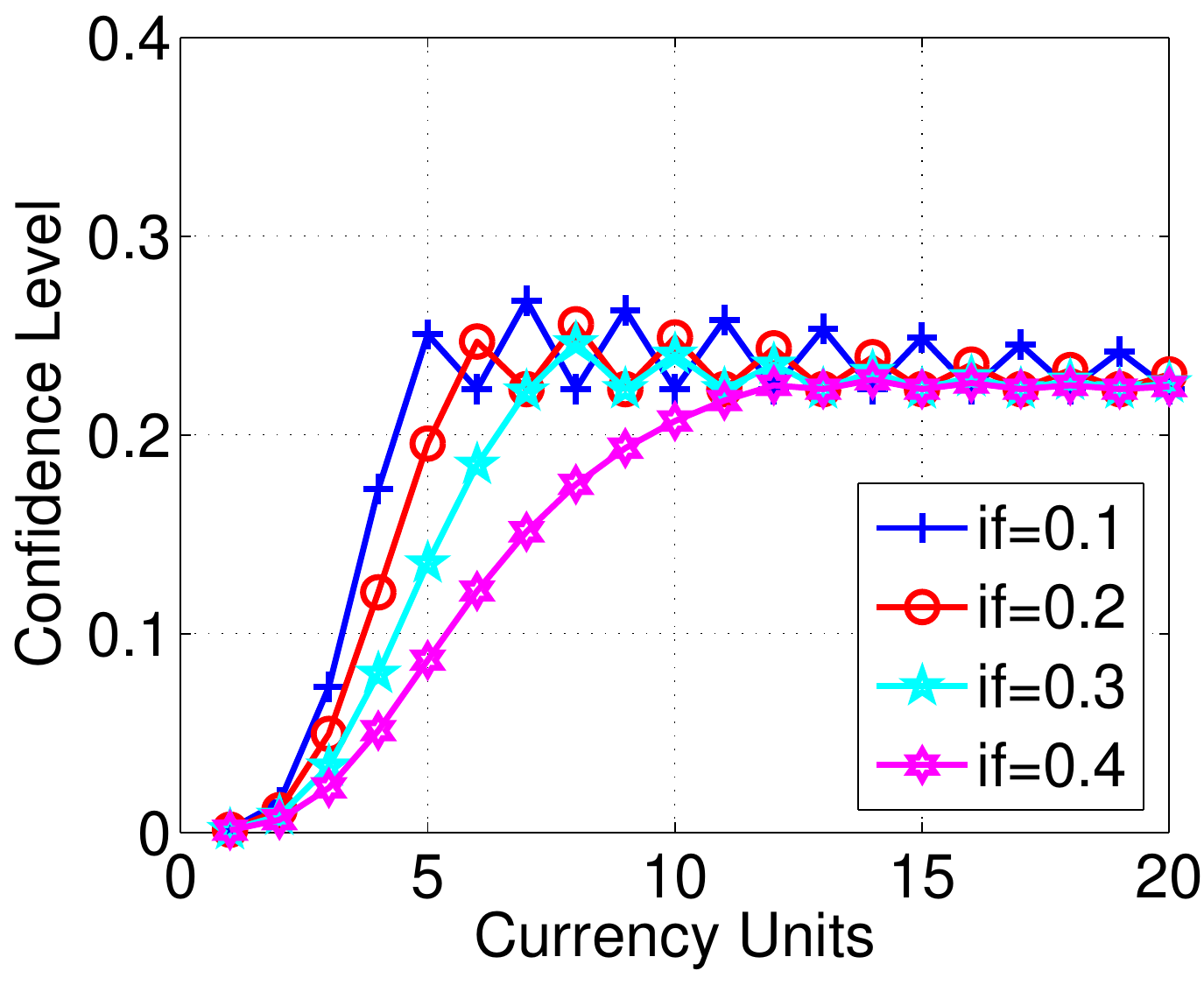}
}
\subfigure[malicious providers $mp=6$] { \label{fig_if:b}
\centering
\includegraphics[width=1.6in]{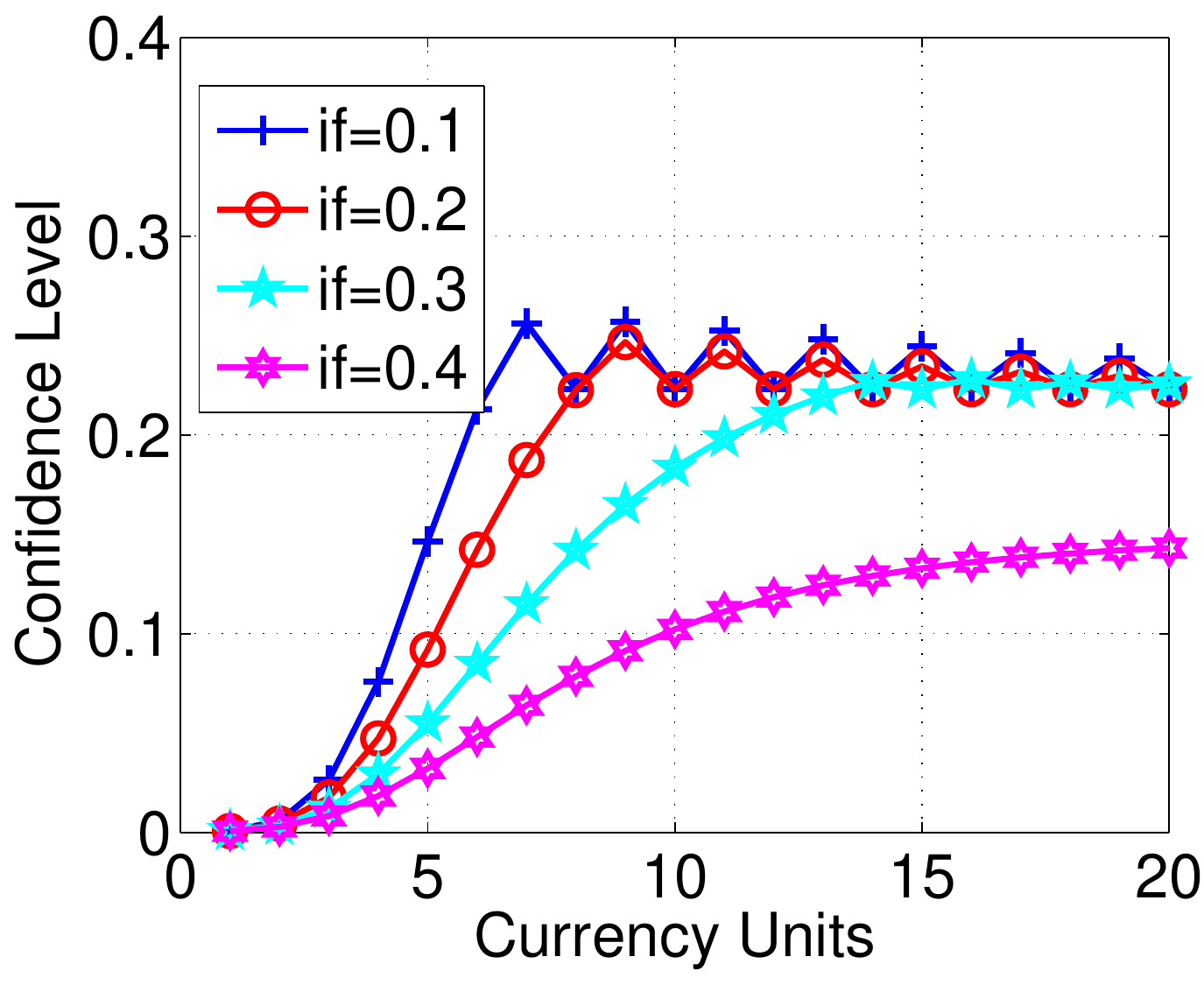}
}
\caption{Confidence level under varying improvement factor $if$}
\label{fig_if}
\end{figure}

We experiment on varying improvement factor $if=0.1$, $0,2$, $0.3$ and $0.4$. The manipulation factor $mf$ is set to $1.6$. The results of malicious providers $mp=3$ and $mp=6$ are shown in Figure \ref{fig_if:a} and Figure \ref{fig_if:b} respectively. The results show that (1) the confidence level grows faster when we have smaller improvement factor; (2) to prompt the confidence to the same level, the cost of currency units is greater in $mv=6$ case compared with in $mv=3$ case. In addition, we can get similar results under other parameter settings.

\section{Related Work} \label{related_work}
IDCS is related to the work on designing digital currency scheme, resolving conflicts from multiple sources, and learning from crowd.

\subsubsection{Designing digital currency scheme} Related studies mainly focus on security issues in scheme designment. E-cash \cite{scheme3} firstly proposes blind digital signatures for trading with electronic currency units. Then later, it is extendedly applied with other considerations, e.g. using RSA digital signatures \cite{scheme4}, constructing group blind signature scheme \cite{group}, and sharing publicly verifiable secret \cite{public}. Also, E-cash scheme provides the ability of fair payment \cite{fair1}\cite{fair2}. Bitcoin \cite{Bitcoin} is the scheme that attracts the most attention recently. It can be implemented with adding features, e.g. Litecoin \cite{Litecoin}, Primecoin \cite{Primecoin}, and Zerocoin \cite{Zerocoin}.

\subsubsection{Resolving conflicts from multiple sources} An early common conflicts resolution method \cite{ra_1}\cite{ra_2}\cite{ra_3} is to average (or to say vote) those conflicts to calculate a truth. However, this type of method suffers from large error when there exist sources with low quality views. To deal with this problem, many methods were proposed to find the truth based on heuristic clues, i.e., prior knowledge on facts \cite{rb_a1}, source dependency \cite{rb_b1}, sensitivity and specificity \cite{rb_c1}. Usually, this type of method uses the clues to evaluate the reliability of sources, and calculate a truth by weighting the sources.

\subsubsection{Learning from crowd} Learning from crowd is another related field to out work. It infers true values from the data labeled by a crowd. The methods \cite{rc_1}\cite{rc_2}\cite{rc_3}\cite{rc_4} proposed in this research field usually focus on specific application scenarios.

\section{Conclusion} \label{conclusion}
We propose IDCS for trading imprecise commodity. It connects buyers and providers through a ledger server, and completes a trade in three stages. We present IDCS prototype implementation on weight assignment, and thus a confidence level can be given for a buyer to decide the quality of a commodity view. In experiment, we characterize the performance of IDCS prototype under varying impact factors.

In this paper, we assume that buyer and ledger server are honest. In the future, we will consider the dishonest case that buyer and ledger server are with malicious intent.

\end{document}